*Review*

# Classification Framework and Chemical Biology of Tetracycline-Structure-Based Drugs

**Domenico Fuoco** [1,2]

[1]  Italian National Board of Chemists and Italian Chemical Society, Rome, 00187, Italy

[2]  McGill Nutrition and Performance Laboratory, Department of Oncology, School of Medicine, McGill University, 5252 Maisonneuve Street, Montreal, QC, H4A3S5, Canada; E-Mail: domenicofuoco@live.ca; Tel.: +1-514-913-1983; Fax: +1-514-504-2077

**Abstract:** By studying the literature about tetracyclines (TCs), it becomes clearly evident that TCs are very dynamic molecules. In some cases, their structure-activity-relationship (SAR) are well known, especially against bacteria, while against other targets, they are virtually unknown. In other diverse fields of research—such as neurology, oncology and virology—the utility and activity of the tetracyclines are being discovered and are also emerging as new technological fronts. The first aim of this paper is to classify the compounds already used in therapy and prepare the schematic structure that includes the next generation of TCs. The second aim of this work is to introduce a new framework for the classification of old and new TCs, using a medicinal chemistry approach to the structure of those drugs. A fully documented Structure-Activity-Relationship (SAR) is presented with the analysis data of antibacterial and nonantibacterial (antifungal, antiviral and anticancer) tetracyclines. The lipophilicity and the conformational interchangeability of the functional groups are employed to develop the rules for TC biological activity.

## 1. Introduction

The number of articles published on tetracycline drugs has reached the threshold of 50,000 papers since 1948. Over the last 10 years, technological fields are emerging in bacteriology and cellular physiology of eukaryotic cells. However, chemical mechanisms of tetracyclines are not completely understood in terms of their function in human cells and, to this day, no (Q)SAR model is validated without doubts. Tetracyclines were first discovered by Dr. Benjamin Dugger of Lederle Laboratories in the mid 1940s as the fermentation product of an unusual golden-colored soil bacterium aptly named *Streptomyces aureofacians* [1]. Tetracyclines (TCs) are a class of antibiotics able to inhibit protein synthesis in gram positive and gram negative bacteria by preventing the attachment of aminoacyl-tRNA to the ribosomal acceptor (A) site [2]. This mechanism has been confirmed by X-ray crystallography [3]. TCs bind specifically to the bacterial ribosome and not specifically with eukaryotic ribosomes. TCs belong to a notable class of biologically active and commercially valuable compounds. This fact may be simply illustrated by mentioning the most important clinical application of TCs, their employment as broad antimicrobial-spectrum antibiotics for human and veterinary use [4]. While they were initially developed as antibiotics, they also hold promise as non-antibiotic compounds for future study and use. Tetracyclines, as dynamics entities, possess unique chemical and biological characteristics that may explain their ability to interact with so many different cellular targets, receptors and cellular properties [5]. The discovery of new uses for tetracyclines and their novel biological properties against both prokaryotes and eukaryotes is currently under investigation by numerous scientists throughout the world. TCs as drugs show only few side effects: one is chelation of calcium and subsequent intercalation in bones and teeth while the other is somewhat like photosensitizing drugs, given their phototoxic action on keratinocytes and fibroblasts. The mechanisms of phototoxicity *in vitro* and *in vivo* are not yet entirely clear [6].

**Figure 1.** Structure-Activity-Relationship (SAR) of Tetracyclines (TCs). Shaded: Contact region with 30S rRNA. In blue polygonal: same anthracycline region.

Upper peripheral modification region

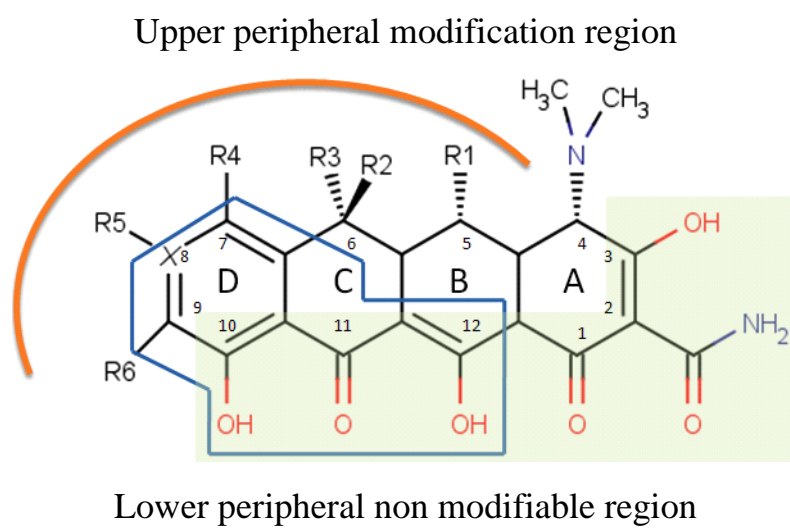

Lower peripheral non modifiable region

## 2. Pharmacological Activities

The therapeutic uses are as follows: antibacterial and non-antibacterial. In the literature, these uses fall into five main categories, namely: (I) newer and more potent tetracyclines used in anitibacterial resistance [7], (II) the nonantibacterial uses of tetracyclines targeted toward inflammation [8] and arthritis [9–12]; (III) in neurology: (a) In tissue destructive diseases acting like antifibrilogenics [13]; (b) Inhibiting caspase-1 and caspase-3 expression in Hungtington's disease [14]; (c) Ischemia [15]; (d) Parkinson's [16] and other neurodegeneration diseases; (IV) antiviral and anticancer [17–19]; (V) Tet repressor controlled gene switch [20].

### 2.1. Antibacterial Use

Currently, as a consequence of their overuse, bacteria have developed TC resistance (efflux pump type) as opposed to the oldest compounds. Medicinal chemists with the intention to optimize structure and improve the antibacterial power have successfully introduced an alkaline group on C-9 of minocycline skeleton, starting as a compound from total synthesis: Tigecycline (a patent of Pfizer and Wyeth, available in therapy from 2005). Searching for new molecules, it is not only important to study the binding of drugs specifically to bacterial ribosomes, but also to understand how the tetracycline skeleton can act as a chelator and ionophore [21]. Moreover, the next generation of antibacterial tetracyclines is currently in progress and will be highly specific for bacterial species and will contain new groups and new rings on the classical skeleton [22]. Mechanism of action of TCs is divided into two categories: "Typical", if they act as bacteriostatic; "atypical", if they act as batericidic. Typical TCs bind specifically to the bacterial ribosomal subunits. All of them that do not have ribosomes as their primary target are considered atypical. Moreover, these atypical mechanisms of action are very toxic both for prokaryotes and eukaryotes (even for mammalian cells). Until now, all TCs used in therapy are broad-spectrum against microbial agents, but researchers are developing a platform to introduce in therapy only novel TCs with a narrow-spectrum for infectious diseases [23].

### 2.2. Non-Antibacterial Use

Both laboratory and clinical studies have investigated the anti-inflammatory properties of tetracyclines. These include: Acting as an inhibitor forlymphocytic proliferation [9], suppression of neutrophilic migration [10], inhibition of phospholipase $A_2$ [11] and accelerated degradation of nitric oxide synthetase [12].

In recent times, starting from the end of the 1990s [24], TCs have showed to be anti-caking of β-amyloid protein and are therefore useful in the treatment of neurodegenerative diseases like Alzheimer's and the Prion Diseases [25,26]. In particular, Minocycline reduces inflammation and protects against focal cerebral ischemia with a wide therapeutic window [27]. Also, Minocycline inhibits caspase-3 expression and delays mortality in a transgenic mouse model of Huntington Disease [28]. Researchers focused on the mechanisms of intracellular pathway communication and genetic control leading to the attenuation of microglia activation [29] and protection of Schwann cells [30].

In the same way that tetracyclines act as pro-apoptotics in neuronal cells, they also act in peripheral metastasis of generalized tumor cells. Experimental data using various carcinoma cell lines and animal carcinogenesis models showed that doxycycline, minocycline and chemical modified tetracyclines (CMTs) inhibit tumor growth by inhibiting matrix metalloproteinase (MMPs) and by having a direct effect on cell proliferation [18,19]. The first use of tetracyclines in viral infections was reported by Lemaitre in 1990 [31]. In 2005, Zink [32] documented the first anti-inflammatory and neuroprotective activity of an antibiotic against a highly pathogenic viral infection. Minocycline is also significantly effective against West Nile virus replication in cultured human neuronal cells and subsequently prevented virus-induced apoptosis [33].

The tetracycline-controllable expression system offers a number of advantages: Strict on/off regulation, high inducibility, short response times, specificity, no interference with the cellular pathway, bioavailability of a non-toxic inducer, and dose dependence. The tet-off system [34], which uses the tetracycline-responsive transcriptional activator (tTA), and the tet-on system [35], which uses the reverse tetracycline-responsive transcriptional activator (rtTA), provides negative and positive control of transgene expression.

## 3. Classification of Tetracyclines

Historically, tetracyclines are considered **First generation** if they are obtained by biosynthesis such as: Tetracycline, Chlortetecycline, Oxytetracycline, Demeclocycline. **Second generation** if they are derivatives of semi-synthesis such as: Doxycycline, Lymecycline, Meclocycline, Methacycline, Minocycline, Rolitetracycline. **Third generation** if they are obtained from total synthesis such as: Tigecycline. However, some researchers consider Tigecycline to be distinct from other tetracyclines drugs and are considered as a new family of antibacterials called Glycylcyclines.

The present paper introduces a new schematic point of view about the denomination and the classification of Tetracycline-Structure-Based drugs. In the years to come, new TCs, which now are advanced in the clinical trials (Phase III of Pharmaceutical Trials Protocol), will be available in therapy. TCs obtained via total synthesis, such as Tigecycline, are considered members of the Third generation if they show wide spectrum activities (both Gram$^+$ and Gram$^-$) Aminomethylcycline derivatives are considered in the same way as Glycylcycline (Scheme 1). In the last five years, thousands of medicinal chemists around the world have synthesized, tested and patented more than 310 tetracycline similar compounds, in particular in the USA [36]. Harvard University [37] and Tetraphase [38] made available pentacycline antibacterials (a structural modification of Doxycycline with five rings), azatetracycline and flurocycline (heteroatoms insert into the D ring, as show in Figure 1) and alkylaminotetracycline antibacterials.

**Scheme 1.** Wide-spectrum of tetracycline activities as antibiotic drugs. TCs are subdivided in: Antibacterial (with typical and atypical mechanism of action), antifungal and antineoplastic.

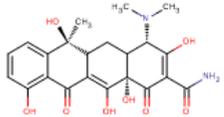

All these compounds are the "logical results of modification around the four rings of tetracyclines that historically started with the master work of Golub and McNamara [39,40] when, thirty years ago, the first eight compounds called chemically modified tetracyclines (CMTs) were introduced in the literature. In 1983, Golub and McNamara [39,40] introduced a new concept concerning the therapeutic usefulness of tetracyclines. They proposed two main ideas. First, tetracyclines, but not other antibiotics, can inhibit the activity of collagenase—a specific collagenolytic metallo-neutral protease produced by host tissues which has repeatedly been implicated in periodontal destruction. Second, this newly discovered property of the drugs could provide a novel approach to the treatment of diseases, such as periodontal diseases, but also certain medical disorders (e.g., non-infected corneal ulcers), which involve excessive collagen destruction. In these cases, TCs appear to inhibit collagenase activity by a mechanism unrelated to the drug's antibacterial efficacy. In fact, all CMTs have been modified by the removal of the dimethylamino group from the C4 position on the A ring. To better classify the new compounds it is very important to understand which chemical properties make it possible for tetracycline-structure-based drugs to act as a "chameleonic" entity. As discussed in the previous paragraph, TCs can be considered as wide-spectrum antibiotics versus bacteria, fungi, virus and cancer cells. In this view TCs can be considered an *optimum* example of multi target drugs and the first well documented in the literature. Moreover, Doxorubicin and all the other anthracyclines are structurally correlated to tetracyclines and it is appropriate to classify them both in the same scheme because of their chemical similarities, chemical physics properties and their use as anti-cancer drugs.

## 4. Chemical Biology of Tetracyclines

### 4.1. Structures Activities Relationship (SAR)

Figure 1 shows the TCs rigid skeleton with the numeration of the four rings, groups and the upper and lower sides of the molecule such as they are commonly called. Many of the chemical modifications of both the first and second generation tetracyclines produced variably active or inactive compounds. An active tetracycline (antibacterial activity) must possess a linearly arranged DCBA naphthacene ring system with an A-ring C1-C3 diketo substructure and an exocyclic C2 carbonyl or amide group. All TCs that act as inhibitor of protein synthesis in bacteria need the amino group in position C4 and keto-enolic tautomers in position C1 and C3 of the A ring. The amino group in the C4 position is pivotal for the antibacterial activities (Scheme 2). A C4-dimethylamino group with its natural 4S isomer is required for optimal antibacterial activity, while epimerization to its 4R isomer decreases Gram–negative activity [41]. It also requires a C10-phenol and C11-C12 keto-enol substructure in conjunction with a 12a-OH group (Scheme 2) outlining a lower peripheral region. All those substituents, with the respective tautomeric equilibrium, are indispensable for recognition and bonding in ribosomal subunits, where chemical modification abolishes bioactivity. Modification of the amide in C2 is possible but with loss of potency. Positions C5 to C9 can be chemically modified to affect their bioactivity and they are designed for the upper peripheral regions, generating derivatives with varying antibacterial activity. Groups R1, R2 and R3 are modifiable to give more selectivity to the biological target in antifungal TCs, but not for the antibacterial activity (Scheme 2). The D ring is the most flexible to change. All modifications of group R4, R5 and R6 are allowed to give highly bacterial specificity and deep changing in pharmacokinetics as result of modifying log P (Table 1).

**Scheme 2.** Structure Activity Relationship of Tetracycline family drugs. Since their introduction in therapy, into the early 1950s, tetracyclines have constantly been modified according to the capabilities of the pharmaceutical laboratories in a given time. Starting as antibacterials, tetracyclines have demonstrated antifungal, antiviral and antitumor properties as well. Nowadays, the new tetracyclines are the most powerful drugs for serious skin infection and the future prospective goal is to separate their anti-inflammatory properties form the antibiotic properties.

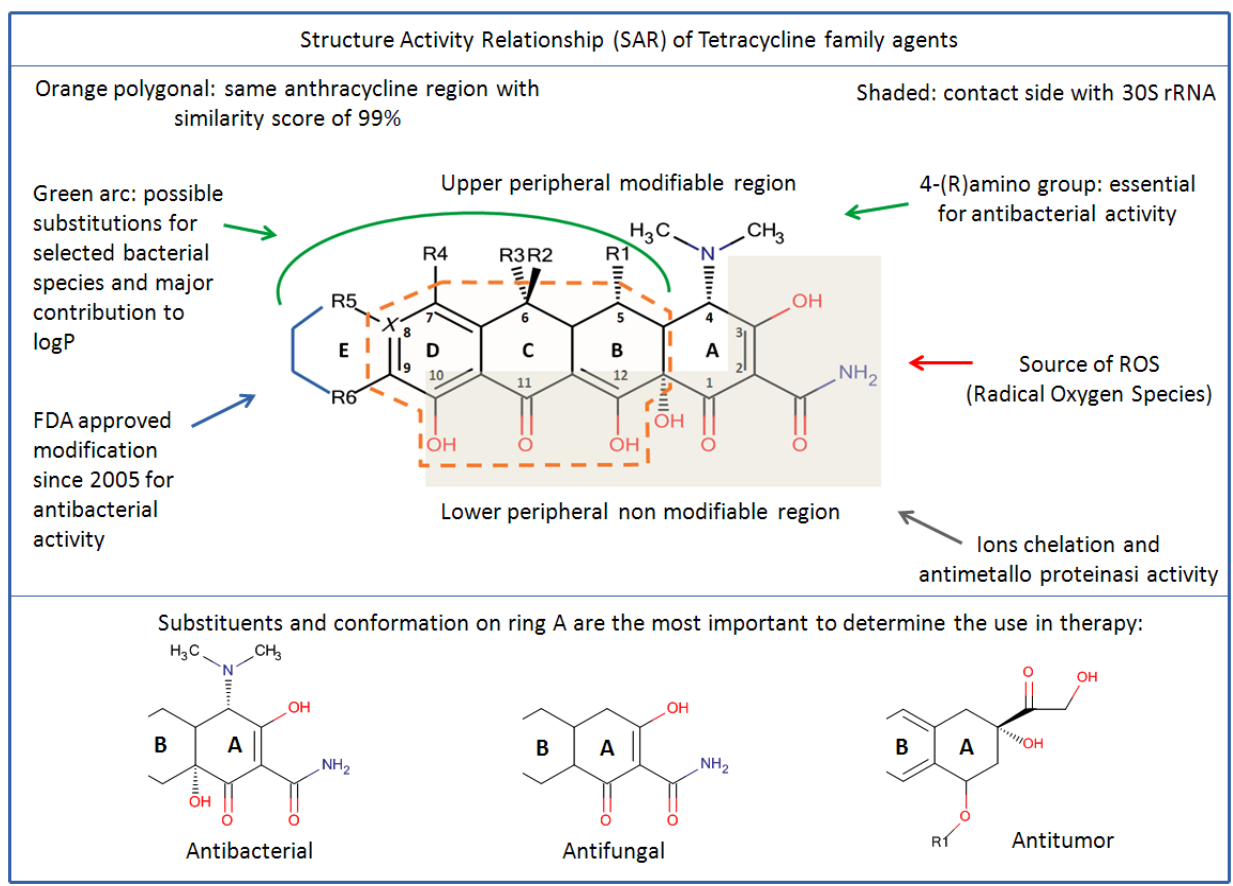

**Table 1.** Experimental data of TCs and their pharmacokinetics values (adapted from www.drugbank.ca).

| Compounds | LogP | LogS | % Enteric absorption | % Serum protein binding | Renal clearance (mL/min) | Half-Life (hours) |
|---|---|---|---|---|---|---|
| Oxytetracycline | −1.3 | −3.14 | 58 | 30 | 90 | 10 |
| Tetracycline | −0.3 | −3.12 | 80 | 60 | 65 | 9 |
| Doxycycline | −0.2 | −2.87 | 93 | 85 | 16 | 15 |
| Demeclocycline | 0.2 | −2.52 | 66 | 75 | 31 | 13 |
| Chlorotetracycline | - | - | 30 | 55 | 35 | 6 |
| Meclocycline | - | - | - | - | - | - |
| Minocycline | 0.5 | - | 95 | 90 | 10 | 20 |
| Rolitetracycline | - | - | - | - | - | - |
| Tigecycline | 0.8 | - | - | 90 | - | 32 |
| Doxorububicine | −0.5 | - | - | 70 | - | 55 |

*4.2. Diverse Mechanism of Action*

Hydroxyl groups are a source of radical oxygen species (ROS) that irreversibly damage macromolecules such as DNA, RNA and proteins. All these effects are considered as fundamental to in cellular death for oxidative stress. In the case of antitumor anthracyclines and CMT-3, these act to block the enzymatic transcriptional complex formation on DNA and then induce apoptotic events. Antifungal and antitumor TCs act in a different way from antibacterial TCs. Once passed into the eukaryotic cells, TCs change the electronic balance equilibrium sequestering divalent ions (e.g., $Ca^{2+}$) much more then monovalent ions (e.g., $K^+$). It is known that TC forms complexes in different positions with calcium and magnesium ions that are available in the blood plasma [42,43]. Most tetracycline acted as bacteriostatic or typical, as protein synthesis inhibitors against bacteria. But it was found that more lipophilic tetracyclines were atypical, with a bactericidal mechanism that relied on membrane damage (as ionophores). Now, medicine is showing the tetracyclines as a family are chemically and biologically dynamic, with multiple mechanisms of activity and capable of interacting with multiple targets, either ribosomal or cellular membranes.

*4.3. Structural Dynamics*

TCs have different acid groups in their structure and the possibility to adopt different conformations. The different proton-donating groups of this molecule offer several possibilities for metal ion substitution. The complexation with metal ions increases the stability of the various TC derivatives. In 1999, Duarte [44] developed a computational and experimental study to evaluate the weight of the various chemical tautomeric behaviors of tetracyclines in solution. The degree of protonation of TC depends upon the specific tautomers that in aqueous solution are more stable than others. It is important to analyze all the possible tautomers of this molecule in their different degrees of protonation and conformations to understand the role of tautomerism in the chemical behavior of TC. Duarte [44] optimized the structures of all 64 tautomers and calculated their heats of formation ($\Delta Hf°$). There are different tautomers in equilibrium in each degree of protonation of TC. They have similar stabilities and they are present in considerable amount in the medium. All the substituent groups contribute with steric effects and polar induced vectors to the geometrical shape of each compound (Figure 2).

## 5. Methodology

In this work a new framework is introduced for the classification of old and novel TCs and their Structure Activity Relationship. The possibility to open access of large chemical data base has changed enormous. All data reporting in this paper has been verified and compared with the National Health Institution Public Library (Bethesda, MD) using PubChem Project. The computational analysis has been performed on a data set of 1325 TCs with a similarity score of 90%, starting from more than 322,000 compounds recorded in PubChem (this paper is in preparation). From that set were chosen the best in class TCs with a similarity score major of 95% (Scheme 1). For each TC selected, PubChem shows at least 112 tautomers and conformers structure with "rule of five" data. The new classification proposed is based on a medicinal chemistry approach due to the complexity of the novel TCs drugs, e.g.,: Chemical modified tetracycline (CMT), Anthracycline drugs (Doxorubicin) and structurally-correlated-tetracycline (Pentacyclin, Glycylcyclin, Flurocyclin, Aminomethylcyclin and Azatetracylin).

**Figure 2.** 3D geometrical shape of Tetracycline (**A**), Tigelcycline (**B**) and SF2575 (**C**). TCs show different conformation and biological activities due their functional groups on the same rigid molecular skeleton.

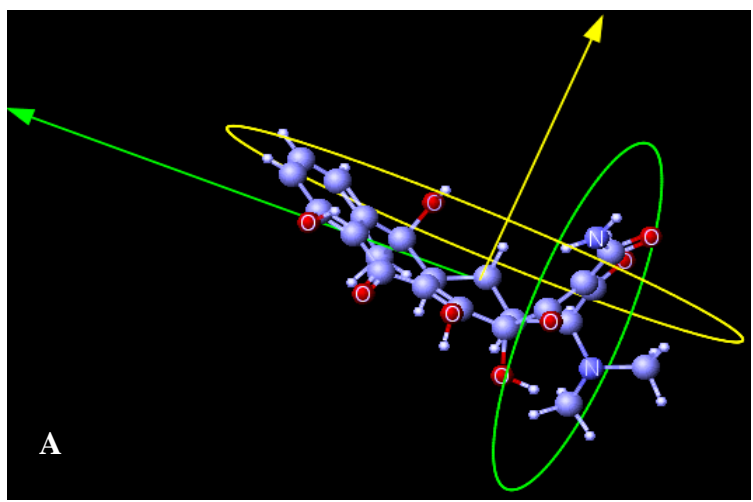

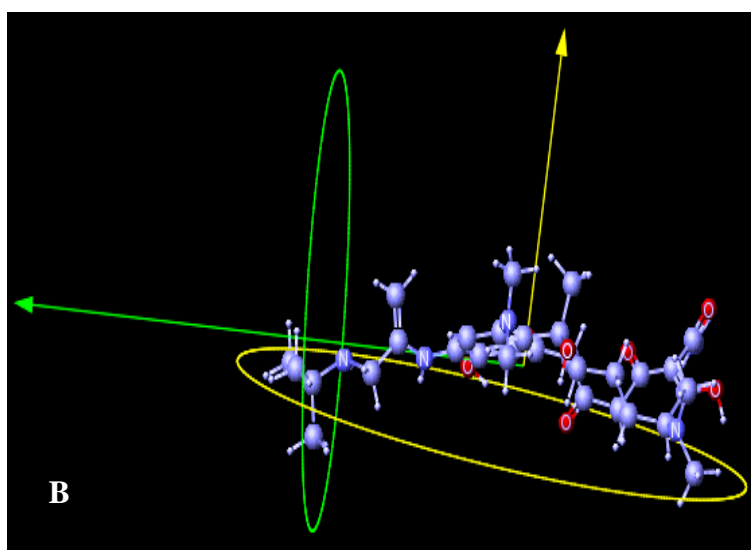

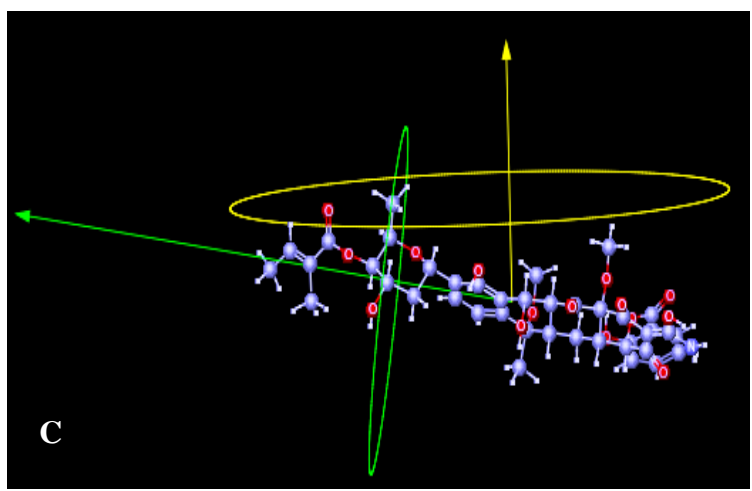

## 6. Conclusions

The role of lipophilicity of TCs is the main factor to explain the biological potency and dynamics of this old family of drugs. It is important to understand the conformational flexibility and the affinity of TCs of metals ions due to the several functional groups present in the molecular skeleton to obtain a significant SAR study. After more than 60 years of scientific investigation, TCs are considered a master example of the pleitropic family of compounds with promising therapeutic properties. Due to the wide use in therapy and their low toxicity, TCs can be considered the first Multi-Target Drugs fully and well discussed in the history of pharmacology.

Newer and more potent antibacterial compounds can be expected to fight the tetracycline-resistant pathogens. New, third-generation derivatives have been designated to be more potent, especially against bacteria possessing ribosomal protection and efflux mechanisms. TCs have increased in potency over time compared to other structural classes of antibiotics. TCs have shown a complex mechanism of action towards various targets due to their conformations under physiological conditions. Such conformations, depending on pH and metal ion concentrations, allow TCs to act in a "chameleonic-like" manner in a large number of diseases. As already suggested, the future of TCs will have increased utility as one of the best anti-inflammatory drugs.

## Acknowledgments

This review is dedicated to my wife and to my family-in-law for their unconditional support to my work. The author wishes to thank Karina Mastronardi for her comments in the editing of this paper.

## Conflict of Interest

The author declares no conflict of interest.